# A Multi-layer hierarchical inter-cloud connectivity model for sequential packet inspection of tenant sessions accessing BI as a service

Hussain Al-Aqrabi, Lu Liu, Richard Hill, Nick Antonopoulos
*Distributed and Intelligent Systems Research Group, School of Computing and Mathematics, University of Derby*
*Derbyshire, United Kingdom*
H.Al-Aqrabi@derby.ac.uk

*Abstract*—Business Intelligence (BI) has gained a new lease of life through Cloud computing as its demand for unlimited hardware and platform resources expandability is fulfilled by the Cloud elasticity features. BI can be seamlessly deployed on the Cloud given that its multilayered model coincides with the Cloud multilayer models. It is considered by many Cloud service providers as one of the prominent applications services on public, outsourced private and outsourced community Clouds. However, in the shared domains of Cloud computing, BI is exposed to security and privacy threats by virtue of exploits, eavesdropping, distributed attacks, malware attacks, and such other known challenges on Cloud computing. Given the multi-layered model of BI and Cloud computing, its protection on Cloud computing needs to be ensured through multilayered controls. In this paper, a multi-layered security and privacy model of BI as a service on Cloud computing is proposed through an algorithm for ensuring multi-level session inspections, and ensuring maximum security controls at all the seven layers, and prevent an attack from occurring. This will not only reduce the risk of security breaches, but allow an organisation time to detect, and respond to an attack. The simulations present the effects of distributed attacks on the BI systems by attackers posing as genuine Cloud tenants. The results reflect how the attackers are blocked by the multilayered security and privacy controls deployed for protecting the BI servers and databases.

*Keywords-component; Business intelligence, Cloud computing, Security and privacy, BI as a service, Cloud tenants vaults, Cloud tenants metadata.*

I. INTRODUCTION

Business intelligence (BI) with big data analysis has emerged as the key future technologies to be hosted on Cloud computing [1]. As per the reports by IBM Tech Trends 2011 and McKinsey Global Institute, these technologies on Clouds are the most sought by business organizations and industries for key emerging business applications like big data analytics, textual analytics, web analytics, network analytics, and mobile data analytics [2][3]. The emerging business areas needing BI with big data are E-commerce, market intelligence, E-government, E-politics, science & technology, E-healthcare, public safety, and public security [2][3][4].

New capabilities of Cloud computing are emerging amidst new expectations and new threats especially because of multi-tenancy environments and emerging security issues in virtualization environment and hypervisor vulnerabilities [5]. BI on Cloud is implemented employing XML data files with DTD (document type definition) structures [6]. The DTD format behaves like a relational database on the Cloud with characteristics similar to OLAP cubes [6]. This format, in a hierarchical data structure, can support heavy OLAP queries and its lightweight architecture can enable it to support big data on the Cloud with faceted search features [7]. BI with OLAP on Cloud can be configured by converting traditional data warehouse tables into XML data files, interlinking the data files using hierarchical DTD structures, defining all data files with the help of suitable structured metadata, and hosting the data files on distributed servers configured to generate a massive parallel processing environment when multi-tenant query loads are applied [6][7]. The computational power of Cloud can be assigned selectively to the OLAP queries through Cloud elasticity (like, Amazon Elastic Compute 2) and the service-oriented platforms (like, Amazon Web Services and Google App Engine) can be used for organizing the XML data files, their DTD structures, and the metadata repositories [8]. The reporting and visualization services may be delivered through dashboard services in BI applications delivered through service-oriented SaaS [8]. The architecture is multi-layered with BI applications, decision support tools, access enablers, data management, data storage, and data sources forming separate layers on the Cloud within the framework of ETL (Extract, Transformation, and Loading) [9]. The Cloud layers can be suitably mapped with the BI layers within the ETL framework [8][9]. The BI data warehouses on the Cloud may be formed by XML data files extracts from traditional self-hosted business transactions databases that could enable planning, forecasting, budgeting, and business control functions through web-enabled and mobile-enabled interfaces [10]. The key business uses could be relational documents access, rich multimedia reports, collaborative information, linking internal and external reports, and data on a click (low-latency data access [10]).

BI on the Cloud requires multi-layered security for ensuring appropriate protection of each BI layer [11]. The key security layers for BI on the Cloud are lightweight directory access protocol (LDAP), intrusion detection and prevention (IDPS), Antispam, web services security, antimalware, and database activity monitoring [11]. The key security services for BI on Cloud are identification, authentication, authorization, auditing, data confidentiality, data integrity, data availability, and prevention from unauthorized DB querying and transactions [12]. This research presents a multi-layer service-oriented security and privacy framework for BI in the Cloud with tenants' session inspection occurring at each layer such that a decision for permitting or denying the session could be made. This architecture is suitable for outsourced private and community

Cloud models [13]. The multi-layer Cloud modeling and the models representing public Cloud, outsourced private Cloud, and outsourced community Cloud are reviewed, and the proposed a multi-layered modeling approach through an algorithm solution for ensuring maximum security controls at all the seven layers are presented in the next section.

## II. CLOUD MODELS FOR THE PROPOSED SOLUTION

Clouds can be modeled in the form of a seven-layer framework, comprising layer 1 as the physical infrastructure components layer, layer 2 as the virtualization resources abstraction layer, layer 3 as virtual resources composition layer, layer 4 as the IaaS, layer 5 as the PaaS, layer 6 as the SaaS providers' applications layer, and layer 7 as the tenants' applications layer [14]. NIST recommended multiple Cloud deployment models incorporating the Cloud layers in the form of five deployment scenarios – public Cloud, onsite private Cloud, outsourced private Cloud, onsite community Cloud, and outsourced community Cloud [13]. The proposed multi-layer service-oriented framework for securing BI on the Cloud is proposed for outsourced private and community Clouds. In the outsourced private Cloud model, multiple corporate owning BI applications, data marts, and data warehouses can host their resources on a Cloud service provider committed to serve organizational clients only [13]. In this model, each corporate client could be provided secured VPN access to the respective BI resources [13]. The outsourced community Cloud model will also have multi-organization BI implementations, with a difference that the organizations can access each other's resources by virtue of community-based agreements [13]. The proposed architecture comprises multi-layered security and privacy services offered at a cost, which may be unaffordable to retail Cloud clients. Hence, public Clouds have been kept outside this solution. In an outsourcing environment, each proposed layer is a Cloud in itself offering the specific security or privacy control in a service-oriented framework. For example, the anti-malware layer offers an array of Cloud databases comprising records of signatures and traces of malware entering through compromised sessions (i.e., embedded in the session packets).

Before getting into the solution, a review of security and privacy as a service on Cloud computing is presented in the next section.

## III. SECURITY AND PRIVACY AS A SERVICE

The security as a service is a multi-layer framework in which, appropriate security and privacy controls are positioned on each layer of the seven-layer model of Cloud computing [15]. In this model, every security layer mapped with the Cloud layer may be offered as a chargeable service to the clients [16]. Such a framework may be deployed as per the principles of trustworthy computing at each layer [16]. Given that IaaS access layer is made of virtual machines, all the security services will be accessible through VMs allocated to the tenant organizations [17]. The virtualization security manager may be embedded within the virtual machines manager that controls application profile allocation to end clients [17].

The security and privacy controls need to be defined at the service interfaces between the tenants and the Cloud [18]. Hence, the key controls need to be positioned at the service virtualization abstraction layer [18]. For privacy controls, each tenant may be assigned a vault comprising privacy resources like digitally signed documents, encryption keys, tokens, and digital signatures [19][20]. This layer may be added over an authentication layer and a tenants' metadata layer defining all attributes of the authorized tenants required for accessing appropriate resources on the Cloud [21][22][23]. A session can be denied by either of these layers if the tenant-specific information in the session packets does not match the database contents serving these layers. This concept is presented in the form of a detailed multi-layer design in the next section.

## IV. THE MULTI-LAYER SECURITY MODEL

The model (Figure 1) is multi-layered Cloud architecture with each layer delivers one of the security/privacy services. The tenant sessions are allowed to pass through an array of firewalls acting as gateways. The firewalls are made of Cisco PIX 535 models acting as network, transport, and application layer firewalls. The firewalls can be equipped with access control lists based on protocols, IP addresses, hostnames, TCP/UDP ports for common applications and services, and higher order TCP/UDP ports for custom applications and services. At the application layer, the firewall comprises filters for allowing URLs, HTTPS sessions, FTP sessions, Java clients, and Active X controls, selectively. In addition, the firewalls can encrypt all sessions using IKE (Internetwork Key Exchange) with IPSec (IP Security) protocol. The encryption services supported are DES (56 bits), 3DES (128 bits), and AES (256 bits).

The tenants are hosted on four tenant LANs accessing the Cloud network through four firewalls. In addition to the tenant LANs, three hackers are shown accessing the Cloud through different firewalls for simulating a distributed attack. The hackers have been modeled as clients gaining network and transport level access to the Cloud by breaking the firewalls. They can also be viewed as valid tenants that have gained access to the Cloud by buying a published subscription.

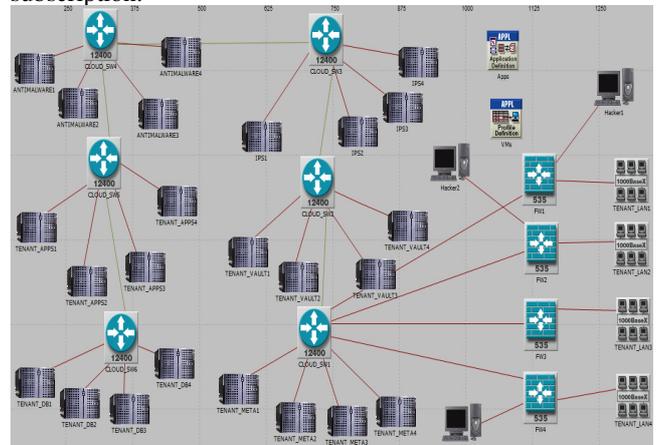

Fig. 1. The model architecture

The Cloud is made of seven layers as shown in Figure 2. Each layer may be viewed as a Cloud in itself. The services of the layers are delivered by arrays of servers deployed in them to ensure that there is no performance bottleneck. The tenant sessions enter the Cloud through the firewalls and pass through the layers in a sequential manner. Each layer verifies the sessions and allows further communications with the next layer above it [24]. After a series of examinations, the sessions finally reach the Cloud apps layer running the BI applications. The roles of the layers are explained as the following:

a) Tenants' firewalls – are deployed for ensuring that authorized tenants are allowed to enter the Cloud; includes built-in authentication using RADIUS, protocol.
b) Tenants' Metadata – comprises detailed tenant information for ensuring their authorization to destination application instances and database objects; permissions details are embedded in the tenant sessions.
c) Tenants' Vaults – comprises decryption keys or digital signatures for unlocking the destination application instances and database objects.
d) Intrusion Prevention System – checks for traces of exploit signatures in the ongoing sessions.
e) Anti-malware – checks for viruses and spyware signatures in the ongoing sessions.
f) Tenant Apps – comprises suites of BI applications.
g) Tenant DB – comprises a large repository of database objects serving the BI applications.

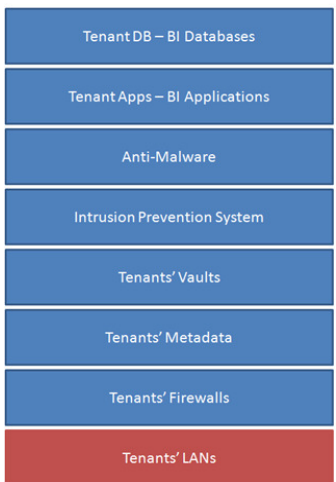

Fig. 2. Cloud layers modeled

The Cloud layer services are modeled as shown in Figure 3. Each database service has a built-in relational database for examining the tenant sessions. Tenant metadata comprises tenants' information for enabling authorization to designated application instances and database objects that the tenant has subscribed. Tenants' vault comprises keys and digital signatures stored in database objects that can be retrieved based on tenant authorization enabled by the metadata layer. The IPS comprises a database of known exploit signatures such that a tenant's attempt to launch exploits through an authorized session can be matched with them, detected, and blocked. The anti-malware comprises a database of known malware traces such that a tenant's attempt to spread malware through authorized sessions can be detected and blocked. The Cloud DB comprises the database objects that the tenant is authorized to access. These objects are locked and can be unlocked by the keys picked up from the tenant vault layer. OPNET does not have the feature to enter content in databases. Hence, the simulations are limited to studying their operating behavior, performance levels, and blocking of attacks.

The applications are packaged in profiles as shown in Figure 3. In this design, the profiles represent a pool of virtual machines in which, all the security/privacy services, the Cloud BI apps, and the Cloud BI database are packaged. There is no profile configured for allowing access to these applications services outside the virtual machines. All the servers on the Cloud run these virtual machines profiles only.

Fig. 3. Virtual Machines with applications packaged

An attack scenario is presented in Figure 4. In this scenario, the attacker may gain access to a VM in the capacity of a verified tenant. The Cloud service providers offer a number of subscriptions and the tenant is given access to their virtualized environments after some preliminary verification (like ID and address proof, bank account details, credit card details, etc.). However, the tenant may be a hacker attacking neighboring VMs through virtual links. As per the scenario shown in Figure 4, the attacker may deploy effective exploitation tools like Metasploit and Netmapper on their VMs. Metasploit works in a terminal emulation mode. The attacker may simply have to store its files on the virtual disks on the Cloud accessible through the VM and launch the terminal emulation. It launches an emulation screen and allows the attacker to execute thousands of exploits from its internal database of exploit codes and their payloads. The entire process is quite user-friendly and expert attackers can even modify its codes to make it operative in any environment. The process is described in detail on metasploit.com. It is also used as a penetration-testing tool given its stealth capabilities. Metasploit can help the attacker to penetrate the neighboring VMs and create covert channels through which, data proliferation can be carried out without anyone knowing about it.

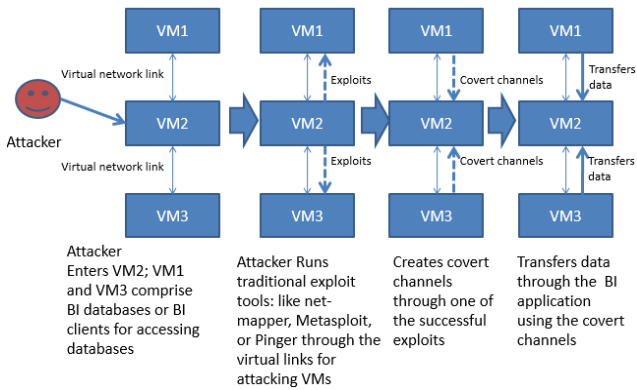

Fig. 4. An attack scenario showing the problem

When such exploits are launched from the VM owned by the attacker, the Cloud security controls will not have any ways to know about the attacks. This is because the attacker has already entered the maize of VMs hosting hundreds and thousands of them. The Cloud providers may be having sound peripheral security controls around this maize of VMs for protecting against external attackers. However, what could be done when an attacker has access to a VM deep inside the maize by merely buying a premium subscription? There needs to be a solution to such a scenario. The solution can be implemented by converting the maize of VMs into a hierarchical framework as presented in Figure 5. In the modified scenario, the attacker is shown as entered VM2 through a formal subscribing process of the Cloud service provider (perhaps, paying the highest subscription and getting a premium user status). The attacker can attack VM1 and VM3 through cross-channel attacks (VM to VM) because the virtual links are insecure. The security controls are normally deployed outside the hosts having VMs and hence they can protect against external attackers only. In this scenario, the attacker has gained access to VM2 and hence is an insider attacker. This might have happened through subscription validations.

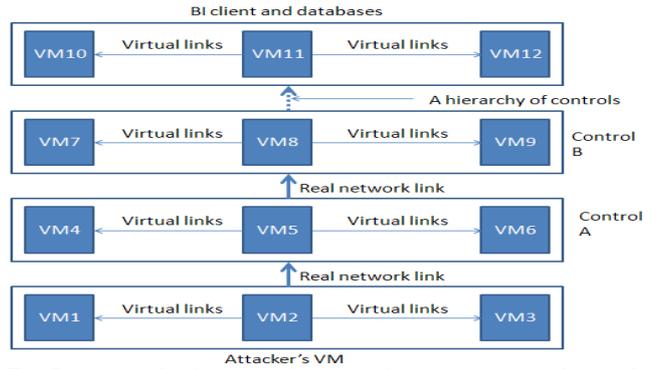

Fig. 5. A hierarchical framework presented as a solution to the attack scenario in Figure 4.

However, in this architecture the attacker will not gain anything by attacking VMs 1 and 3 because they have nothing except a BI client that can proceed only through the next control (example, Tenant Metadata inspection). Hence, the attacker will have no choice but to proceed through the Real network link and establish session with VMs 4 to 6. These VMs hold the first control (tenant metadata inspection). The attacker may cross this control if he/she has genuine tenant credentials. However, the attacker will be countered by the second control in VMs 7, 8, 9. In this way, the attacker will have to breach all control layers successfully before reach the VMs hosting the BI databases. In order to breach the databases for stealing data of other tenants, the attacker will have to use exploit tools (like, Metasploit, NetMapper or Pinger). However, these exploits will be detected by the IPS and Anti-malware layers. Hence, the attacker may fail to steal any data in spite of gaining access to Cloud VMs by buying subscriptions. For example, this design can protect Amazon EC2 Cloud that has been tested to be vulnerable to VM to VM cross channel attacks.

Figure 6 presents the mechanism in which the controls may be deployed. The sequencing of controls may merely be security policy decision for making the system as effective as possible. The concept will remain the same irrespective of how the controls are organized. Figure 6 shows the path of the session of a tenant travelling through multiple VMs before reaching the BI application and its databases.

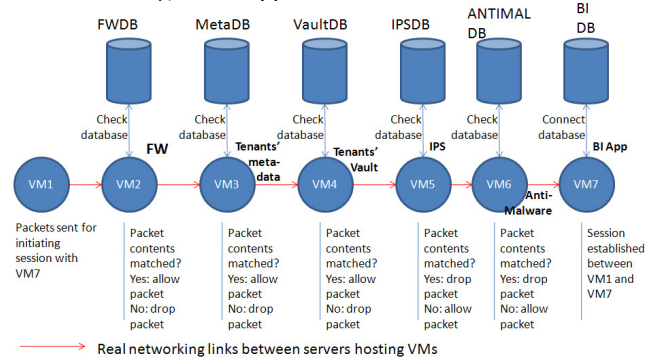

Fig. 6. The controls positioned on each layer of the hierarchy

The VM1 comprises a lightweight client of the BI application loaded in VM7. The client knows nothing except that it needs to connect to VM7 to proceed further. Hence, an attacker in an adjacent VM will gain nothing by attacking this VM. The attacker will have this client anyways on his/her native VM. The client may be viewed as an empty window in which, the subsequent controls will launch screens for capturing details. At VM2, the client session will be tested for the source VM (by virtue of a VM identification code) and a screen will be launched for entering authentication details. At VM3, the client will launch a screen for capturing the tenant details. It may be in a form for randomly asked questions and answers. For example, the form may ask to enter mobile number, social security code, and zip code to proceed further. The questions may vary in each new session. The tenant cannot move further without answering the questions. The VM2 will check the answers in MetaDB and allow the session to proceed. At this stage, the tenant cannot forge answers unless an exploit is used to penetrate the MetaDB and steal information about other tenants. The MetaDB is expected to be secured and patched with latest updates. However, if the attacker is successful the session can proceed to the next level. It may be noted that the exploit traces have entered the session packets and hence attacker will have no means to clean them before proceeding. This is because the attacker cannot interrupt the session and inspect the packets of the ongoing session. A session interrupter needs to be placed between the VMs 4 and 5, which is not there by design. These VMs security controls accessible only to the security administration team of the Cloud service provider. Hence, assuming that the attacker is able to breach VM4 to steal digital signatures or private keys, as well, the traces of exploit will be detected at VMs 5 and 6. Both the layers are equipped with databases of all known exploit and malware signatures. Hence, the sessions cannot escape them. The only scenario of their success can be the zero day attack when a new form of exploit or malware has been used to penetrate the VMs 3 and 4 and the databases of VMs 5 and 6 are not updated with their information. Now-a-days, security vendors are working actively to counter zero day attacks through their IPS or other form of solutions. A study of websites of security companies like Symantec, Cisco, Trend Micro, etc. will reveal about their research and new innovations. These companies are actively designing their solutions for virtualization platforms. This architecture is not about the capabilities of IPS and antimalware controls albeit is about how these controls can be deployed in a multi-layer hierarchy. In fact, any form of new controls can be added to this solution by simply adding a Cloud layer. Once the session screen passes all the controls, it reaches the BI App VM where the screen for launching reports and dashboards will be displayed. Given that all these layers will have multiple VMs with parallel computing, the inspections can be completed within a few seconds thus ensuring acceptable performance. However, it needs to be considered that the multi-level inspections of session packets will induce delays. While the VMs 2, 3, and 4 will not inspect the packets of an allowed session, the IPS and antimalware in VMs 5 and 6 will continue to inspect each packet passing through them.

All these controls will be software-based controls with active databases supporting them. They are expected to act at the platform-as-a-service layer except the firewall, which is expected to act at the infrastructure-as-a-service layer. The mapping of these controls with the seven-layer Cloud model presented by references [14] and [15] is presented in Figure 7. The Figure 7 shows that the layers 1, 2, and 3 will comprise tenant VMs with BI clients and no data stored. The sessions from tenant VMs will go through multiple validations in layers 4 and 5 before reaching layer 6 for accessing the SaaS BI application or layer 7 for accessing tenants' customized applications and databases. In this way, the proposed architecture is positioned on layer 4 (IaaS) and layer 5 (PaaS). The firewall is treated as an IaaS control because it will authorize sessions by inspecting VM IDs and hence has a link with the virtualization and composition layers. The VM IDs will be assigned automatically at layers 2 and 3 and the access controls for them will be assigned at the layer 4 in the firewall. The remaining controls are not concerned with the VM identification because they are deployed for packet inspections. Hence, they may be perceived as platform controls. There may be application-level controls as well, which are not shown in this architecture. For example, the BI apps may have own authentication layer. In addition, the tenants may have an additional layer of authentication layer established for accessing data based on their organizational roles. For example, the CEOs may have different access levels than the employees.

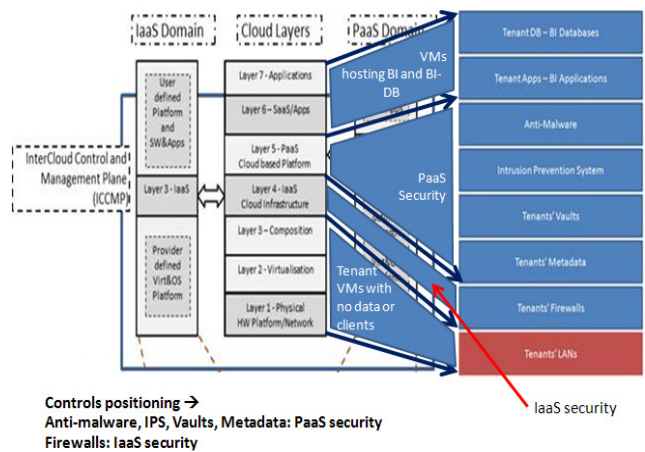

Fig. 7. Mapping of the proposed architecture with the seven-layer Cloud model

The proposed solution is presented in the form of the following algorithm. The algorithm can be modified if the security policy of sequencing the controls is changed. However, the basic structure of the algorithm will remain the same.

S = tenant session
$P_{ct}$ = contents of session packets;
$DB_{fw}$ = contents of FW
$DB_{meta}$ = contents of TENANT_META;
$DB_{vault}$ = contents of TENANT_VAULT;
$DB_{ips}$ = contents of IPS;
$DB_{antimal}$ = contents of ANTIMALWARE;

S = 1 If $P_{ct} \in \{DB_{fw}, DB_{meta}, DB_{vault}\}$ AND $P_{ct} \notin \{DB_{ips}, DB_{antimal}\}$);
= 0 Otherwise;
Here, 1 = permit, and 0 = deny.
{
*Step 1: Initiate S;*
*Step 2: Do steps 3-19;*
*Step 3: Set S = 1, Match $P_{ct}$ ;*
*Step 4: If $P_{ct} \neq DB_{fw}$; S = 0, Go to Step 5*
   *Else, go to Step 6*
*Step 5: End IF;*
*Step 6: Set S = 1, Match $P_{ct}$*
*Step 7: If $P_{ct} \neq DB_{meta}$; S = 0, Go to Step 8*
   *Else, go to Step 9*
*Step 8: End IF;*
*Step 9: Set S = 1, Match $P_{ct}$*
*Step 10: If $P_{ct} \neq DB_{vault}$; S = 0; Go to Step 11*
   *Else, go to Step 12*
*Step 11: End IF;*
*Step 12: Set S = 1, Match $P_{ct}$*
*Step 13: If $P_{ct} = DB_{ips}$; S = 0; Go to Step 14*
   *Else, go to Step 15*
*Step 14: End IF;*
*Step 15: Set S = 1, Match $P_{ct}$*
*Step 16: If $P_{ct} = DB_{antimal}$; S = 0; Go to Step 17*
   *Else, go to Step 18*
*Step 17: End IF;*
*Step 18: Set S = 1;*
*Step 19: Loop*
}

The contents of the session packets ($P_{ct}$) are matched with the contents of the databases positioned at each checkpoint ($DB_{fw}$, $DB_{meta}$, $DB_{vault}$, $DB_{ips}$, and $DB_{antimal}$). The session is initiated by the client machine (through a browser based VM) at Step 1. The first level of inspection is at Step 3. The session should contain VM ID provided by the Cloud service provider. There should be an entry of the VM ID in the firewall database ($DB_{fw}$), else the session will be dropped. The next checkpoint is at step 6. The session should comprise authentication-related information (like, userID token, password token, unique client ID, and such other identifiers assigned by the Cloud administrators). The authentication information contained in the session packets should match the registries in the tenant metadata ($DB_{meta}$). The session is allowed only if the information contained in the session packets match the registry entries. The next checkpoint is at Step 9 for verifying if the session with its unique VM ID and authentication ID has been allocated a private key to access encrypted database rows. The session will be dropped if there are no decryption keys assigned to the designated unique VM ID or unique tenant ID. If the session passes this level, it is a fully authenticated and authorized session. However, in this design the packet inspectors for Exploits, Trojan codes, and malware are positioned after full authentication and authorization. This has been done to protect the Cloud from insider attackers. The inspections and matching with intrusion prevention database ($DB_{ips}$), and antimalware database ($DB_{antimal}$) can block attacks by authorized VMs, as well. The steps 12 and 15 are designed for this purpose.

A closer look at the algorithm reveals that the tenant session initiates in the first hop and scrutinized at each layer. All layers are deployed as separate Clouds. The fundamental rule states that in order to permit the session until the Cloud apps, the packet contents of the tenant session should match the contents of firewalls, tenant metadata, and tenant vault, and should not match the contents of IPS and antimalware databases. Hence, the algorithm can prevent insider attackers, and can enhance multi-tenancy security in the Cloud IaaS, and Cloud PaaS models as well.

Table 1 presents the profile and destination preferences of Tenants' LAN 1. This LAN has 500 client machines and three VMs are allocated for each client named – VM1, VM2, and VM3. It may be noted that the destination settings on this LAN is the Tenants' Metadata servers, and not the main BI applications and databases. This has been configured to ensure that the tenant sessions do not bypass the first security/privacy layer of the Cloud. In real networks, this control can be implemented either in the applications manager of the virtual network controller or in an application layer firewall. Similar settings have been made in the destination preferences of Tenants' metadata servers that point towards the tenant vaults. These configurations justify how the algorithm is supposed to work. They enforce sequential checks of each session packet not allowing them to jump any checkpoint.

Table 1 Tenants' LAN attributes

| Attribute | Value |
|---|---|
| ⊟ Application: Destination Preferences | (...) |
| – Number of Rows | 1 |
| ⊟ TENANT_META | |
| – Application | TENANT_META |
| – Symbolic Name | Database Server |
| ⊟ Actual Name | (...) |
| – Number of Rows | 4 |
| ⊟ Enterprise Network.TENANT... | |
| – Name | Enterprise Network.TENANT_META1 |
| – Selection Weight | 10 |
| ⊟ Enterprise Network.TENANT... | |
| – Name | Enterprise Network.TENANT_META2 |
| – Selection Weight | 10 |
| ⊟ Enterprise Network.TENANT... | |
| – Name | Enterprise Network.TENANT_META3 |
| – Selection Weight | 10 |
| ⊟ Enterprise Network.TENANT... | |
| – Name | Enterprise Network.TENANT_META4 |
| – Selection Weight | 10 |
| ⊞ Application: Source Preferences | None |
| ⊟ Application: Supported Profiles | (...) |
| – Number of Rows | 3 |
| ⊞ VM1 | ... |
| ⊞ VM2 | ... |
| ⊞ VM3 | ... |
| – Application: Supported Services | None |

The Cloud is multi-layered and the destination preferences of one layer are to the subsequent layer only. Hence, a session cannot be established directly with the Cloud apps servers by passing these layers. The content session packets are compared with the database contents of each layer. Based on the fundamental rule, the session either is permitted to the next layer or is denied. For example, if the session contents do not comprise a tenant key identical to the one stored in the vault, it is not assigned an encryption key, and the packets are dropped, based on a rule recommended by [19]. The hackers have been configured as clients having access though the first layer of firewalls because they are the valid tenants (by purchasing an initial subscription on the Cloud). The results of the simulation are presented and discussed in the next section.

## V. ANALYSIS OF RESULTS

Table 2 presents the sessions encountered by the tenant LAN. It may be observed that all the sessions initiated from the tenant LAN are with tenant Metadata only through all the virtual machines. Similarly, the sessions are between tenant Metadata and tenant Vault only for the same virtual machines. This has been observed for all the hops designed in the model. The results indicate that the virtual machines cannot jump a layer given their pre-defined destination preferences. In this way, the session inspections and forwarding/dropping are mandatorily implied on each VM.

Table 2. Client DB sessions on Tenants' LAN

The VMs assigned to the hackers are kept out of the tenant Metadata and the tenant Vault application profiles. In practice, this scenario may be viewed as the unauthorized users not having any records in the metadata or the vault when trying to access a different domain than allowed to them. It may be observed in Figure 8 that the IP packets from the hackers' LAN dropped after an initial attempt. To get through all the layers and establish unauthorized sessions with the Cloud apps, the hackers will need to break the metadata layer, the vaults layer, the IPS layer, and the antimalware layer. It is unlikely that the hackers will be able to break so many security/privacy layers to reach the Cloud apps.

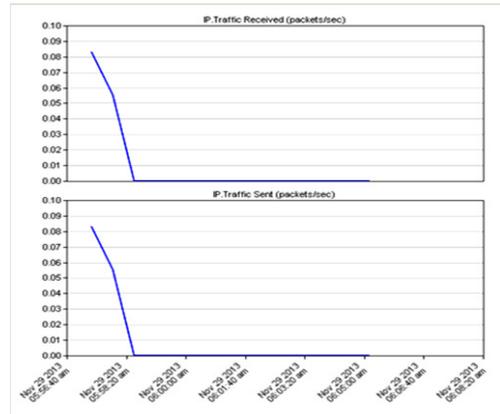

Fig. 8. After initial attempts, IP packets from the hackers' machines are dropped

On the other hand, the authorized tenants' LANs could run DB sessions throughout the simulation period as shown in Figure 11. This is because their VMs are added to the application profiles of all the layers (i.e., their sessions fulfill the rule shown in the algorithm). This is a multi-layer security architecture in which, each layer can be a Cloud in itself and served by a different Cloud service provider. Hence, this entire architecture is termed as a multi-layer security as a service framework for tenant organizations hosting BI applications and databases on outsourced private and community Clouds. These services may be premium and expensive and hence, may not be suitable for public Clouds. In addition, there are challenges of maintaining five inspection-aiding databases for keeping them highly performing and up-to-date. There may be costs charged to tenants for records maintenance in tenants' firewalls, tenants' metadata registries, and tenants' vaults. All access permissions may be based on the Unique VM IDs of the browser-based VM access allocated to the client. The IPS and antimalware databases can be updated regularly through their websites of original software manufacturers (OSMs). Given the volumes expected in these Clouds, such updating will require inter-Cloud communications such that all databases in the IPS and Antimalware arrays can pull records from OSMs databases directly. All the five databases should be hosted on server arrays with parallel processing, as configured in the model. The number of servers per array will be in hundreds and not the few shown in the model. Hence, this model is designed to establish the concept only and not demonstrate the expected volumes.

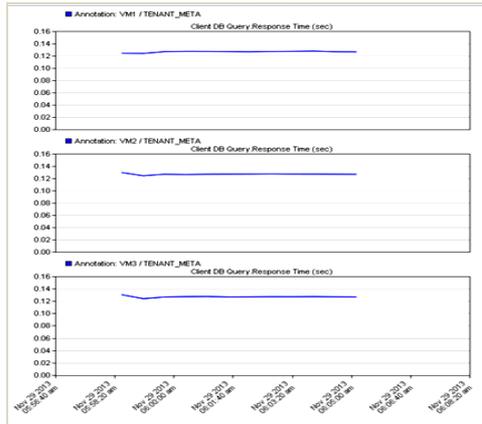

Fig. 9. Authorized tenant LANs established and ran DB sessions with the TENANT METADATA

## VI. CONCLUSIONS

In this paper, a multi-layer model of security and privacy on Cloud computing is proposed for securing BI and its data warehouses/data marts on Cloud computing. The design is based on a multi-step algorithm comprising inspection of session packets at each layer of the design. The contents of the session packets should match the firewall, tenant metadata, and tenant vault layers and should not match the IPS and antimalware layers. The layers are configured in such a manner that the devices of one layer prefer the next layer for its sessions. Given that all sessions are running in the VMs, they need to be scrutinized at all these layers before allowed to the final Cloud apps layer. The hackers could not breach the layers because their VMs were included in the profiles of the firewall layer only. This scenario is similar to a tenant having metadata and vault access attributes assigned by the Cloud service provider, which are unknown to the hacker. Hence, penetrating the firewalls' layer could not serve the purpose of the hackers. However, maintaining this system will require costly services thus making it unsuitable for public Clouds. The solution is mapped with 7-layer Cloud model.